# Hierarchical structural control of visual properties in self-assembled photonic-plasmonic pigments

Natalie Koay[1,2], Ian B. Burgess[2,*], Theresa M. Kay[2], Bryan A. Nerger[1,3], Malaika Miles-Rossouw[1,4], Tanya Shirman[1], Thy L. Vu[1,2], Grant England[1], Katherine R. Phillips[5], Stefanie Utech[1], Nicolas Vogel[1], Mathias Kolle[1,6], and Joanna Aizenberg[1,2,5,7,*]

[1]*School of Engineering and Applied Sciences, Harvard University, Cambridge, MA 02138, USA*
[2]*Wyss Institute for Biologically Inspired Engineering, Harvard University, Cambridge, MA 02138, USA*
[3]*Chemical Engineering Program, University of Waterloo, Waterloo, ON, Canada*
[4]*Chemical Engineering Program, University of Ottawa, Ottawa, ON, Canada*
[5]*Department of Chemistry & Chemical Biology, Harvard University, Cambridge, MA 02138, USA*
[6]*Department of Mechanical Engineering, Massachusetts Institute of Technology, Cambridge, MA 02139, USA*
[7]*Kavli Institute for Bionano Science and Technology, Harvard University, Cambridge, MA 02138, USA*
[*]*ibburges@fas.harvard.edu*, *jaiz@seas.harvard.edu*

**Abstract:** We present a simple one-pot co-assembly method for the synthesis of hierarchically structured pigment particles consisting of silica inverse-opal bricks that are doped with plasmonic absorbers. We study the interplay between the plasmonic and photonic resonances and their effect on the visual appearance of macroscopic collections of photonic bricks that are distributed in randomized orientations. Manipulating the pore geometry tunes the wavelength- and angle-dependence of the scattering profile, which can be engineered to produce angle-dependent Bragg resonances that can either enhance or contrast with the color produced by the plasmonic absorber. By controlling the overall dimensions of the photonic bricks and their aspect ratios, their preferential alignment can either be encouraged or suppressed. This causes the Bragg resonance to appear either as uniform color travel in the former case or as sparse iridescent sparkle in the later case. By manipulating the surface chemistry of these photonic bricks, which introduces a fourth length-scale of independent tuning into our design, we can further engineer interactions between liquids and the pores. This allows the structural color to be maintained in oil-based formulations, and enables the creation of dynamic liquid-responsive images from the pigment.

## 1. Introduction

The use of colorants is ubiquitous in everyday life with applications ranging from food and clothing to cosmetics and paints.[1-4] The fine tuning of hue and effects (such as brightness, texture, sparkle, color travel, etc.) generated by a pigment is contingent upon the ability to control its scattering and absorption profiles. Absorption is most commonly manipulated using organic dyes, which provide the widest variety of available colors, but come with several drawbacks. In addition to having limited long-term photo- and chemical stability, molecular dyes often require a combination of different compounds to achieve different hues. The resulting variety of chemical compounds associated with a broad color palette increases production costs and requires the toxicity of many compounds to be analyzed for the many applications where colorants come into contact with the body (*e.g.* food, clothing, and cosmetics).[3]

Engineering structural colors, originating from wavelength-selective scattering rather than absorption, has long been used in conjunction with molecular absorbers to produce visual effects in pigments (*e.g.* sparkle, color travel).[4-14] Structural color is achieved by controlling the shape, layering, roughness or porosity of materials on length scales ranging from hundreds of nanometers up to millimeters.[12,15-19] While the manufacturing of versatile structural color materials requires exquisite control of fabrication parameters on multiple length scales, it is decoupled from the specifics of the material chemistry or composition. Thus, in contrast to molecular dyes, a multitude of structural color effects can be generated from a single inert material.[4-14]

A class of pigments in which absorption and scattering can be controlled within a single inert, photo-stable, and non-toxic material framework is highly attractive. The development of this prospect has recently spurred a significant amount of research into using structure to control both scattering and absorption, with the goal of eliminating the need for organic dyes with the associated toxic and photobleaching effects. Controlling absorption in addition to scattering is important, because in the complete absence of an absorber (e.g. in purely structural pigments), all incident wavelengths must scatter in some direction. Under diffuse lighting conditions, this generally restricts the accessible purity of colors. While it has been shown that wavelength-selective, but angle-independent scattering can be generated structurally without the need for an absorber,[8] non-scattered wavelengths still exit the structure via transmission in this regime, making the apparent color highly sensitive to the properties of the substrate.

To achieve full control of a pigment's appearance, independent on what it is deposited, one must therefore be able to engineer some form of wavelength-selective light absorption into the design. Keeping the original motivation in mind, the challenge becomes how to be able to widely control absorption within a chemically invariant inorganic framework. One solution that has been explored recently is the incorporation of a single broadband absorber (e.g. carbon black) either inside of[5,9,10,13,14,20,21] or underneath[22] a structural-color material. In this case, the scattering profile of the structural features allows certain wavelengths to interact more with the underlying absorber than others, extracting spectrally selective absorption from a broadband absorber. Since the absorber has a broadband absorption profile, the same absorber can be used for each different color. However, while structure can impart spectral selectivity to the absorption from a broadband absorber, some absorption will still occur at all wavelengths, leading to a trade-off between the brightness and saturation of a pigment's hue (with highly saturated colors appearing dim).

Plasmonics offers an alternative approach to engineering absorption within an invariant chemical composition. Plasmonic absorption resonances, analogous to their photonic counterparts, can be engineered by manipulating the size and shape of metal structures or nanoparticles.[23-31] Additionally, the relevant material length-scale on which plasmonic resonances are engineered is typically much smaller than the corresponding scale for photonic resonances. This provides the opportunity for plasmonic particles to be incorporated inside of traditional structural color materials without disrupting their design. Several groups have recently synthesized 3D photonic crystals doped with plasmonic particles[12,32-36] and it has been shown that they produce composite color profiles with complex interplay between photonic and plasmonic resonances with different wavelength and angular dependences.[33]

Here we describe a simple, one-pot templated co-assembly method for bottom-up synthesis of hierarchical pigment particles consisting of silica inverse-opal bricks that are doped with plasmonic absorbers. We study the effect of interplay between the plasmonic absorbers and the Bragg resonance on the visual appearance of these photonic bricks when they are dispersed with a distribution of orientations on a surface. We then show how manipulating the order of the inverse-opal porosity and the overall dimensions of the bricks allow us to control the prominence of the Bragg resonance as well as the degree to which the bricks can be preferentially oriented. These manipulations let us engineer color effects (e.g. sparkle, color travel) that can either enhance or contrast with the plasmonic absorption resonance. Adding chemical functionalization to the pore surfaces, addressing features at a length-scale below those that were used to manipulate color and effects, we can control the wetting characteristics and the appearance in liquids. This allows the pigment particles to retain bright color in oil-based paint formulations and enables the creation of dynamic images whose appearance changes when wet.

## 2. Synthesis of photonic bricks

Our fabrication protocol is depicted in Figure 1A,B (see Appendix A for details). The photonic bricks are formed via a three-phase evaporative co-assembly process that deposits a silica inverse opal film (IOF) doped with plasmonic absorbers on a photoresist template, as depicted in Fig. 1A. The substrate is vertically suspended in an evaporating solution containing an acidic solution of tetraethyl orthosilicate (TEOS), monodispersed polymer colloids (polymethylmethacrylate or polystyrene, $d \sim 200\text{-}400$nm), and plasmonic absorbers (gold or silver nanoparticles, $d \sim 10\text{-}50$ nm).[33,37-38] The plasmonic absorbers were all coated with a passivation layer (silica or polyethylene glycol) to prevent aggregation (see Appendix A). During evaporation, a convective assembly process leads to the formation of a colloidal crystal in the channels provided by the template, with a silica sol-gel matrix doped with plasmonic absorbers co-assembling into the interstitial sites.[33] As has been previously shown, this type of co-assembly process produces inverse-opal films that are highly ordered, consisting of a face-centered-cubic lattice with a single crystallographic orientation.[37] This order is maintained even when nanoscale plasmonic absorbers are included in the assembly (see Fig. 1C), provided that their dimensions are sufficiently small.[33] This single crystallographic orientation is also preserved when the IOF is deposited into vertical channels as is done here, since the template geometry does not disrupt lattice formation along the growth direction.[38]

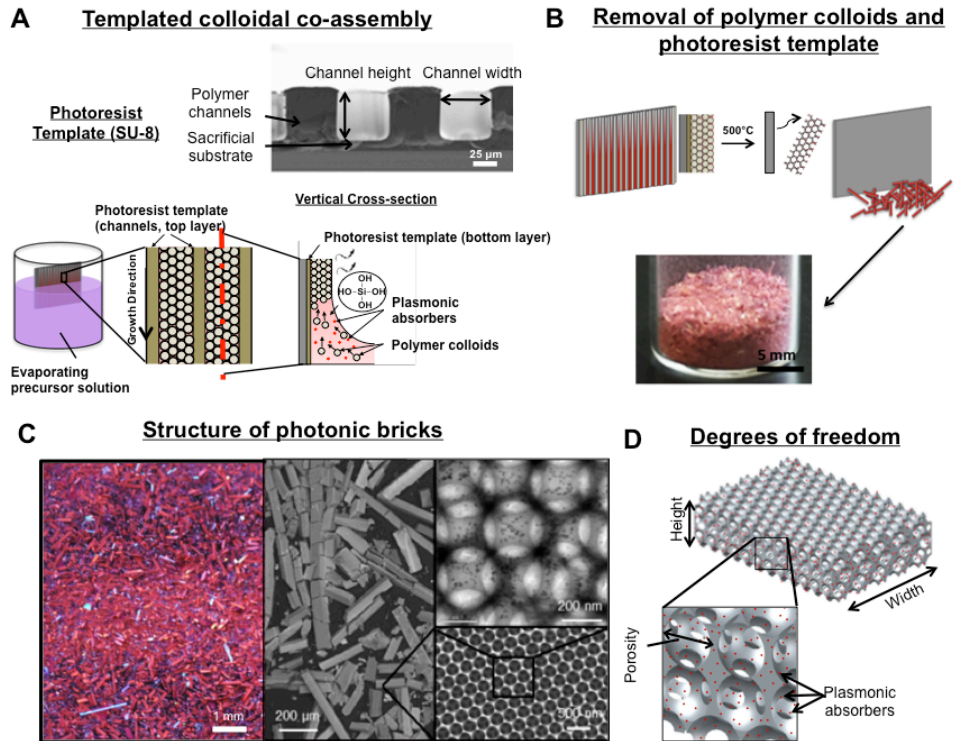

Fig. 1. (A,B) Fabrication protocol: (A) Upper image: Substrates are coated with a template of photoresist (SU-8), comprised of a uniform bottom layer topologically patterned with channels aligned with the growth direction. Lower schematic: Silica inverse-opal films, doped with plasmonic absorbers and plugged with a template of polymer colloids, are deposited in the channels via an evaporative co-assembly process developed recently, see Ref. 33. (B) The photoresist template and the polymer colloids are then simultaneously removed by calcination at 500°C in air, releasing the photonic bricks from the substrate. (C,D) Structure of the photonic bricks: (C) Left: mm-scale optical image (black background, diffuse illumination) of a collection of photonic bricks (photoresist template channel dimensions: 25 μm x 100 μm ($h$ x $w$)) with several insets at different scales. Center: low-magnification SEM image showing the overall shape of the photonic bricks. Bottom right: higher magnification SEM image showing the sub-micron scale porosity. Top right: TEM image showing the gold nanoparticles encapsulated in the matrix. (D) The fabrication protocol allows for tuning of the morphology and composition of photonic bricks on several disparate length-scales (*e.g.* overall dimensions, porosity, plasmonic absorber doping).

The photoresist template consists of topologically patterned polymer channels on top of a flat sacrificial polymer substrate. Both layers are fabricated from SU-8 photoresist using photolithography. This template is used to systematically direct the breakup of the IOF into specific shapes (directed by the shape of the channels[38]) and release them from the substrate (upon removal of the unpatterned base layer), as shown in Fig. 1B, allowing us to control two of the dimensions of the photonic bricks (see Fig. 1D). After a single deposition step (Fig. 1A), the polymer colloids and the photoresist template are removed via high-temperature calcination (Fig. 1B). This simultaneously opens up the porosity by removing the polymer colloids and releases the photonic bricks from the photoresist template. Substrates were calcined inside of glass vials to collect the photonic bricks once they fell from the substrate. The photonic bricks were mounted for imaging and spectral characterization by coating a substrate with tape and then gently patting the substrate onto the inside of the vials containing the photonic bricks until the surface was covered.

The channels in the photoresist template allow us to control two dimensions of the photonic bricks without disrupting the underlying symmetry of the pores.[38] These dimensions, hereafter referred to as the width and height of the photonic bricks, as denoted in Fig. 1D, correspond to the channel width and height (or depth), respectively, in the photoresist template. We used three different channel dimensions of photoresist templates ($h$ x $w$): 25 μm x 50 μm, 50 μm x 50 μm, and 25 μm x 100 μm. The distribution of dimensions of the photonic bricks produced from each of the photoresist templates are shown in Fig. 2. The heights and widths of the photonic bricks were found to be 5-10% smaller than the corresponding template dimensions, likely a consequence of the matrix shrinking as a result of polycondensation during the high-temperature calcination step.[39] There was also slightly larger variability in the heights ($<\sigma_h> = 19\%$) as compared to the widths ($<\sigma_w> = 6\%$), reflecting the higher sensitivity of this dimension to the local concentration of precursors and shape of the meniscus. The length of the photonic bricks (the dimension left unconstrained by the channels so as not to disrupt formation of the colloidal crystal along the growth direction) is determined by cracking of the long strips of IOF that occurs as the photoresist template is removed. The length of the photonic bricks typically followed wide distributions, ranging ~10-1000 μm ($<\sigma_l> = 45\%$).

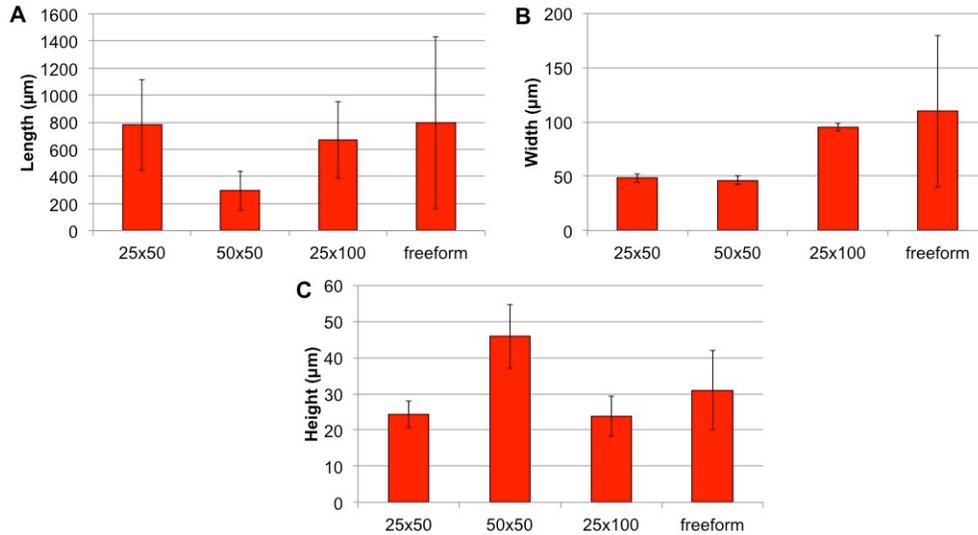

Fig. 2. (A) Length, (B) width, and (C) height distributions of photonic bricks synthesized in photoresist templates with different channel dimensions (standard deviation shown as the error bar), as well as for freeform photonic bricks. The lengths are the only dimension not specified in the templated photonic bricks and are determined by natural cracking of the inverse-opal strips as they are released from the channels

We also grew photonic bricks in the absence of channels, leaving both the length and width of the bricks to be determined by how the IOF cracks as it is released from the substrate. The height can still be tuned in this geometry by adjusting the concentration of the precursor solution. In this case, where we were not looking for fine control of the width and height, we found that breakup and delamination during calcination of IOFs grown on the inner walls of the 20 mL vials used to house the co-assembly solution actually produces larger quantities of these "freeform" photonic bricks than were obtained from an unpatterned photoresist template. Freeform photonic bricks are typically larger in width and similar in height compared to those generated from templates ($<l> \sim 800$ μm, $<w> \sim 110$ μm, $<h> \sim 31$ μm). Broader size distributions were observed for all dimensions of freeform photonic bricks

($\sigma_l$ = 79%, $\sigma_w$ = 64%, $\sigma_h$ = 35%), similar to the size distribution for undirected length of "templated" photonic bricks.

We are able to engineer different aspects of the overall optical appearance through independent adjustments to the morphology of the plasmonic absorbers, the wavelength-scale porosity and the overall dimension of the photonic bricks (Fig. 1D). In the sections that follow, we will study how each morphological tuning parameter affects the optical properties.

### 3. Nanoscale plasmonic absorbers

The nanoscale metal particles doped into the matrix of the photonic bricks provide spectrally selective absorption that can be tuned by altering their size and shape.[14] This type of structural absorber enables us to choose from a variety of underlying absorption profiles to add to the system without needing to change the chemical composition or the pore structure of the photonic bricks. As an illustration of this freedom, we use two different types of plasmonic absorbers here: 1) gold nanospheres (AuNS, $d$ ~ 12 nm, visible in the TEM image in Fig. 1C), which have an absorption maximum at $\lambda$ ~ 520 nm and provide photonic bricks a red hue (Fig. 3A); 2) silver nanoplates, having an absorption maximum at $\lambda$ ~ 650 nm and provide photonic bricks with a blue hue (Fig. 3B). We synthesized the AuNS[40] and passivated them with a thin layer of polyethylene glycol (PEG) to prevent aggregation. Unlike traditional organic dyes, AuNS do not exhibit photobleaching after prolonged light exposure, as shown in Fig. 4 that compares photostability of inverse opal films containing AuNS and those containing an azo-benzene dye of the same color.[41] Although blue hues can also be engineered in gold particles with the correct morphology,[14] we used silver nanoplates for illustrative purposes here due to their commercial availability. However unlike the AuNS, the silver nanoplates did not show the same degree of long-term stability within the photonic bricks, degrading over the scale of weeks at when exposed to the atmosphere at room temperature.

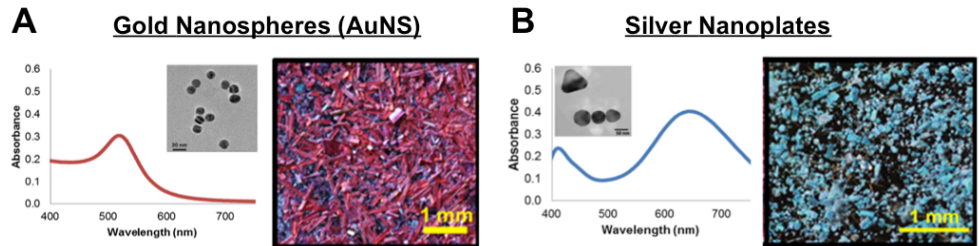

Fig. 3. Absorption is controlled by doping photonic bricks with metal nanoparticles. The absorption profile can be tuned by adjusting the size and shape of the particles. For example: (A) gold nanospheres ($d$ ~ 12 nm) display an absorption maximum at $\lambda$ ~ 520 nm and impart a red hue when incorporated into photonic bricks (0.7% AuNS by solid volume, $a$ = 350 nm, formed from a photoresist template with 25 μm x 100 μm channels), while (B) silver nanoplates (thickness ~ 10 nm, d ~ 60-80 nm), display a primary absorption maximum at $\lambda$ ~ 650 nm and impart a blue hue when incorporated into photonic bricks (0.35% Ag by solid volume, freeform, $a$ = 200 nm).

The plasmonic absorbers provide the photonic bricks with vibrant hues, even when viewed under diffuse lighting conditions. Fig. 5A,B and C (left column) compare the appearance of an inverse opal film (IOF) and bricks, lacking any added absorbers, when viewed at different angles under diffuse illumination. The IOF displays a bright structural color that is due to strong specular reflections in which there is a Bragg resonance. The IOF (having a pore diameter $a$ ~ 350 nm) shown in Fig. 5A appears red when viewed at normal incidence. Although it is illuminated from multiple directions, the single well-defined crystallographic orientation across the film[37] allows only one scattering plane to contribute significantly to the film's appearance. This ordering also causes the appearance of the film to blueshift in color at higher angles of viewing (see Fig. 5A). However, when an IOF is broken up into bricks, each brick now has several additional exposed surfaces and is free to orient at a different angle. This orientational distribution (illustrated schematically in Fig. 5B), along with increased non-resonant scattering from higher surface area, gives the photonic bricks

(with pore diameter, $a \sim 350$ nm) a white appearance under diffuse illumination (Fig. 5C, left column). The addition of plasmonic absorbers counters this loss of hue, which is usually a characteristic of structural color pigments in the absence of absorbers.[5] Unlike the Bragg resonance, the absorption of AuNS is not intrinsically angle-dependent.[33] When AuNS are added to the photonic bricks, blue and green wavelengths are suppressed for all illumination and viewing angles. As a result, the photonic bricks appear red from all angles under diffuse illumination conditions when AuNS are added, and this red hue intensifies with increasing AuNS concentration (Fig. 5C).

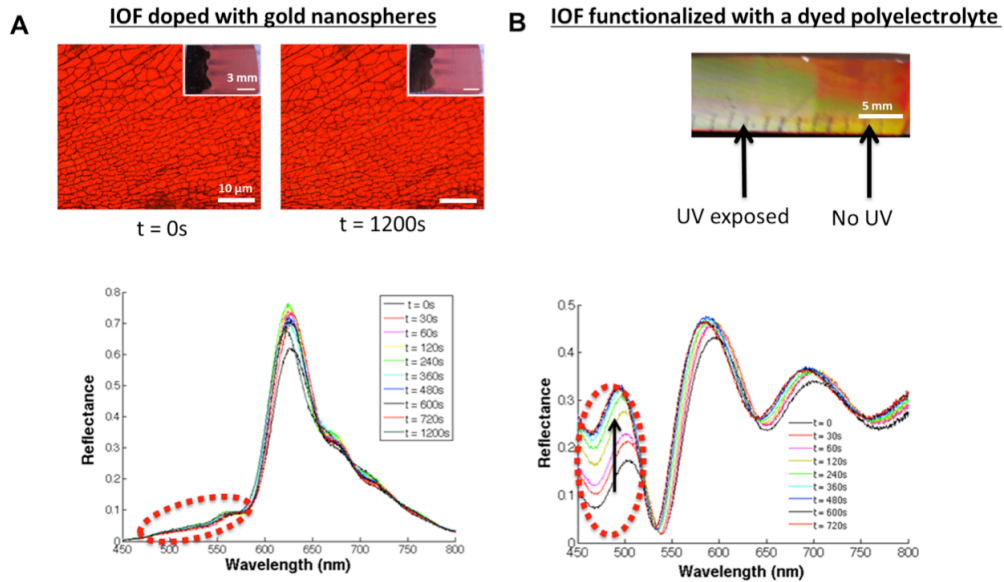

Fig. 4. Comparison of the photostability of an inverse-opal film (IOF) doped with AuNS (0.7% AuNS by solid volume) (A) and an IOF whose pore surfaces were functionalized with a dyed polyelectrolyte (B, poly-(Disperse Red 1 acrylate-*co*-acrylic acid)[41]). While the dyed IOF shows photobleaching after 30 s of UV lamp exposure (130 mW/cm$^2$), no changes in color or spectral response were observed for up to 1200 s of exposure (exposure beyond 1200 s not tested).

The role of the absorber in AuNS-containing photonic bricks is quantified as a function of the AuNS concentration in Fig. 5D,E for photonic bricks that originated from a photoresist template with 25x100 μm channel dimensions. Reflectance from collections of photonic bricks deposited on a substrate was measured as a function of illumination and collection angles using a home-built variable-angle spectrometer.[39] As can be seen in the raw spectra shown in Fig. 5D for AuNS contents of 0%, 0.7%, 1.3% (by matrix volume), the AuNS provide an absorption profile that is angle-independent, even reducing scattering in the Bragg resonance maximum as it intersects the AuNS absorption maximum, which occurs at high incidence angles of around $\theta \sim 40°$. The suppression of scattering both on (red data points) and off (black data points) the Bragg resonance is quantified in Fig. 5E that compares the relative strength of scattering in green (absorbed by AuNS) and red (not absorbed by AuNS) wavelengths. For both, selective suppression of green light scattering initially drops before saturating with increasing AuNS concentration. At our highest level of AuNS doping (2.4% Au by solid volume), there was a $44 \pm 20\%$ selective reduction in scattering of green light on the Bragg resonance compared with an $82 \pm 8\%$ selective reduction of the Bragg resonance. The comparatively weaker effect of the absorber on the Bragg resonance is expected since scattering is stronger, and thus light at resonant wavelengths interacts with the absorbing structure over effectively shorter distances.[5,14]

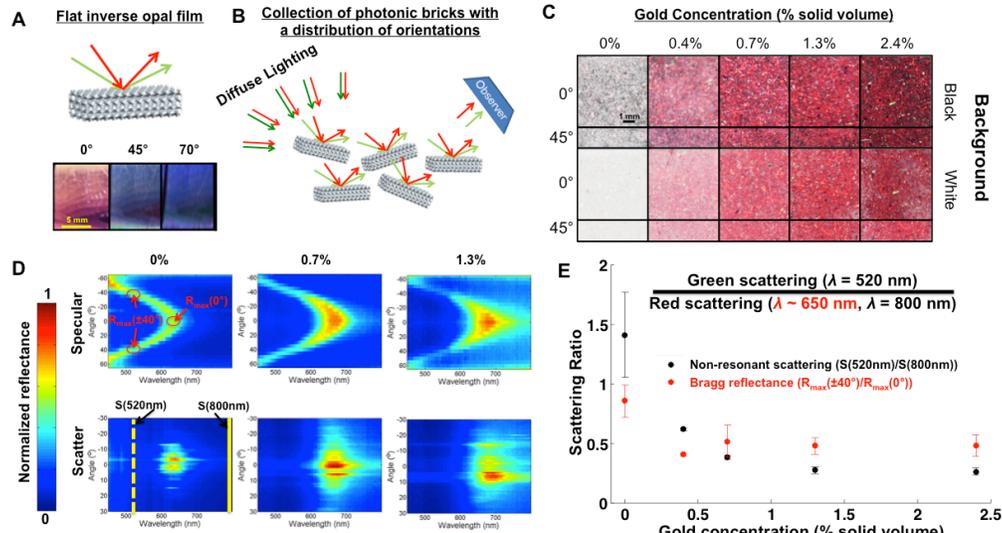

Fig. 5. (A) Inverse-opal films (ordered porosity with $a \sim 350$ nm) display a bright red color in specular reflection, but this color blueshifts with increasing viewing angle. (B) When a single, flat inverse-opal film is replaced by randomly oriented photonic bricks, many scattering angles contribute to the observed color under diffuse lighting conditions, leading to a white appearance (C, left column). (C-E) Doping the photonic bricks (same porosity as A) with AuNS, which absorb blue and green wavelengths, produces a red hue. As a result, dispersions of AuNS-containing photonic bricks appear red. (C) Photographs (diffuse lighting conditions) showing how the red hue of the photonic bricks intensifies with increasing AuNS concentration (all bricks were built from photoresist templates with channel dimensions: 25 μm x 100 μm ($h$ x $w$)). (D) Normalized specular reflectance (top row) and diffuse scattering (bottom row, 0° illumination) of the photonic bricks shown in C) with AuNS concentrations of 0%, 0.7% and 1.3% by solid volume. (E) Suppression by the AuNS ($\lambda_{abs} \sim 520$ nm) of Bragg reflection (red points, $R_{max}(\pm 40°)/R_{max}(0°)$) and non-resonant scattering (black points, S(520nm)/S(800nm)). Suppression of Bragg reflection (red points) is the ratio of the peak Bragg reflectance at $\theta = \pm 40°$ (where $\lambda_{max} \sim 520$ nm) and at $\theta = 0°$ ($\lambda \sim 650$ nm). Suppression of non-resonant scattering (black points) is the ratio of scattering intensity at $\lambda = 520$ nm and $\lambda = 800$ nm averaged over all scattering angles from -30° to 30° (0° illumination). Neither $\lambda = 520$ nm nor $\lambda = 800$ nm overlaps with the Bragg resonance at any of these angles. Both measures are averaged over three samples of photonic bricks for each AuNS concentration.

## 4. Influence of the wavelength-scale porosity and overall dimensions of photonic bricks on visual effects

The Bragg resonance exhibited by photonic bricks with ordered porosity provides strong iridescence. This iridescence, when combined with the hue produced by the plasmonic absorbers, allows us to create complex appearances and effects. One of the simplest effects a Bragg resonance can have is to enhance the brightness and saturation of the hue produced by the plasmonic absorber. This is accomplished by selecting the pore size such that the Bragg resonance peak lies in a spectral region outside of the absorption of the plasmonic absorber.[33] The AuNS-containing photonic bricks shown in Fig. 5 provide an example of this effect. Selecting a pore diameter of $a = 350$ nm results in a Bragg reflection maximum in red wavelengths (650 nm) at normal incidence, while the AuNS absorbs strongly at green and blue wavelengths (Fig 2A).

The combination of a spectrally selective plasmonic absorber and a Bragg resonance produces a more efficient color enhancement than the combination of a broadband absorber and a Bragg resonance, which is more commonly used in structural color pigments.[5,9,10,13-14,20-21] This comparison is illustrated in Fig. 6. The normal-incidence reflectance spectra of analogous IOFs ($a = 350$ nm) with and without AuNS doping are compared in Fig. 6A. In the AuNS-IOF, while significant suppression of the reflectance is observed in the blue and green spectral regions (blue arrow), no reduction of the Bragg resonance at 650 nm is observed (red arrow). This effect is in contrast to broadband absorbers that suppress reflectance at the peak Bragg resonance wavelength at 650 nm in addition to the blue and green wavelengths, leading to duller colors. Fig. 6B shows the simulated reflectance at normal incidence of an IOF doped with AuNS compared with that from an IOF doped with an ideal broadband absorber.

Reflectance is simulated using a transfer matrix model, where the wavelength-dependent optical constants ($n$ and $k$) of the silica-AuNS composite matrix material is calculated using effective medium theory.[42] For doping with an ideal broadband absorber (e.g. carbon black), a wavelength-independent attenuation constant ($k$) is given to the IOF matrix material. While both types of absorbers increase the saturation of the red color by suppressing reflectance in the blue and green spectral region (blue arrow), the broadband absorber also suppresses the Bragg reflection (red arrow). Thus, for a given level of saturation (e.g. ratio of the peak Bragg reflectance to the baseline reflectance), a more intense color is achieved with the spectrally selective plasmonic absorbers.

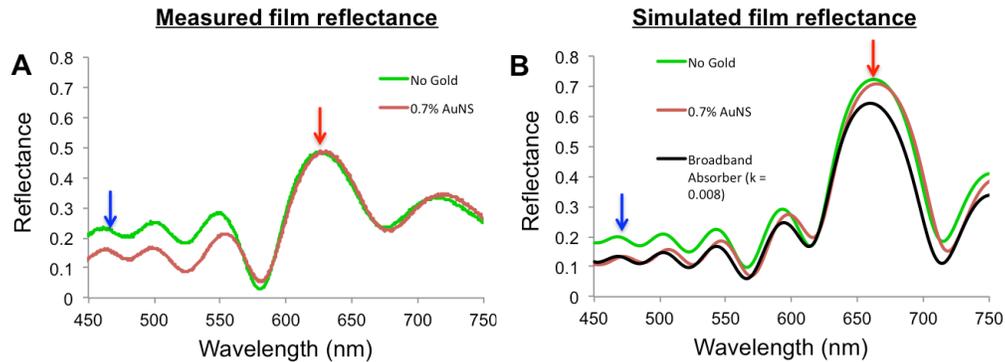

Fig. 6. (A) Measured reflectance spectra at normal incidence for inverse opal films, each with a thickness of 10 close-packed layers on a silicon substrate, with no gold doping (green line) and 0.7% AuNS (by matrix volume). (B) Comparison of theoretically calculated reflectance for inverse opal films (10 layers on a silicon wafer) without AuNS doping, with 0.7% AuNS doping (by matrix volume), and doped with a broadband absorber with a wavelength-independent attenuation constant $k$ =0.008 (chosen to give similar blue-green attenuation to the AuNS-containing film). Reflectance was estimated using a 1D transfer matrix model, where the profile of the porosity is approximated by laterally averaging the dielectric constant as a function of depth. The effect of AuNS doping on the frequency-dependent dielectric constant of the silica was calculated using effective medium theory, as described in Ref. 42.

The Bragg resonance also improves color purity by strongly scattering light over a narrower wavelength range compared to the range of all wavelengths not absorbed by the AuNS. The spectrally narrower the Bragg resonance, the higher the purity of the iridescent colors will be. The width of the Bragg resonance is determined by the refractive index contrast between the air pores and the matrix as well as the length over which the periodicity of the air pores is maintained. A smaller refractive-index contrast leads to a sharper resonance, but requires a larger number of lattice periods to reach a minimum peak width. One advantage of this synthetic protocol is that it enables the production of photonic bricks with ordered porosity and dimensions that are sufficiently large to ensure that the Bragg resonance is as narrow as possible. For this silica:air inverse-opal structure ($n_1$:$n_2$ ~1.46:1.0), a minimum Bragg-resonance width of ~50 nm requires a minimum thickness of ~15 layers or ~4.5 μm (for $a$ ~ 350 nm) to be attained, a length that is significantly smaller than any dimensions of the photonic bricks, both freeform and from the photoresist templates. As can be seen for example in Fig. 5D, the width of the Bragg resonance does indeed approach this limit. In contrast, in many structural color pigments previously studied, such as colloidal glasses, structural coherence is maintained over much shorter length-scales in order to produce an angle-independent hue with a broadband absorber.[5,8,9,10,13-14,20-21] While producing the desired angle-independence of the color, this method also leads to very broad scattering resonances and thus poor color purity.

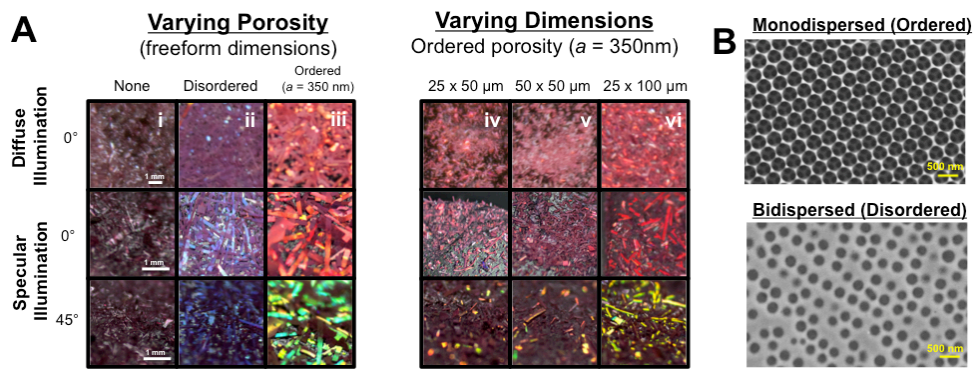

Fig. 7. A) Photographs of photonic bricks dispersed on a substrate with different geometries, all containing the same AuNS content (0.7% by solid volume). The camera's exposure settings were the same for all images in each row to facilitate an objective comparison. The photographs were taken under diffuse illumination (top row, imaged at normal incidence) and with specular illumination at 0° (middle row) and 45° (bottom row). B) SEM images comparing the pore structure of photonic bricks with monodispersed (ordered, top) porosity with $a = 350$ nm and bidispersed (disordered, bottom) porosity with $a_1 = 350$ nm, $a_2 = 240$ nm.

While a Bragg resonance can be used to simply enhance the brightness of the hue of a film at normal incidence, its angle-dependence makes the interplay between the effects of the porosity and the plasmonic absorber more complex in collections of photonic bricks. By tuning their porosity as well as their overall size and shape, we can precisely engineer effects into the pigment's appearance, having the ability to independently address brightness, iridescence, color travel and texture. For photonic bricks with a fixed AuNS content (0.7% by matrix volume), Fig. 7A illustrates the variety of effects that can be produced by manipulating their dimensions and porosity. The images compare the appearance of the photonic bricks with different dimensions and porosity under different lighting conditions (diffuse vs. specular illumination). The left panel (i-iii) compares freeform photonic bricks with three different types of porosity: no porosity (i, synthesized from a precursor solution lacking polymer colloids), disordered porosity (ii, synthesized using bidispersed polymer colloids, $a_1 = 350$ nm, $a_2 = 240$ nm), and ordered porosity (iii, synthesized using monodispersed polymer colloids, $a = 350$ nm). SEM images comparing the ordered and disordered pore structures are shown in Fig. 7B. Since the non-porous photonic bricks (Fig. 7A-i) scatter very weakly, they produce only a dim, angle-independent red hue from the effect of AuNS absorption alone. Disordered porosity (Fig. 7A-ii) enhances the brightness of the hue produced by the plasmonic absorber by increasing the strength of scattering, but also maintains the angle-independence of the appearance. Ordered porosity (Fig. 7A-iii), with the Bragg resonance near 650 nm at normal incidence, provides a much stronger enhancement of the red hue at normal incidence. However, since the Bragg resonance depends on viewing angle, it also introduces a prominent color travel effect when viewed under specular illumination. At glancing angles (e.g. Fig. 7A-iii, bottom row, 45°) the iridescence blueshifts and provides a bright green color that contrasts with the red hue from the AuNS, which can still be seen in the photonic bricks whose orientation does not permit them to reflect the light source specularly to the camera.

The photonic bricks shown in the right panel of Fig. 7A (iv-vi) have the same ordered porosity ($a = 350$ nm), but all exhibit varying degrees of color travel. They are distinct from each other and the freeform photonic bricks (Fig. 7A-iii) only by their width and height that are controlled using photoresist templates. While the appearance under diffuse illumination does not drastically differ between the four (Fig. 7A-iii-vi), color travel is much more prominent in the freeform photonic bricks (Fig. 7A-iii) and those made from the 25 μm x 100 μm photoresist template channels, compared to those made from the two narrower channels (iv: 25 μm x 50 μm, v: 50 μm x 50 μm). In these narrower bricks, iridescence appears more as dispersed iridescent sparkle rather than uniform color travel.

Both the iridescent sparkle of the narrower photonic bricks (25 µm x 50 µm and 50 µm x 50 µm) and color travel of the wider ones (25 µm x 100 µm and freeform: <w> ~ 110 µm and <h> ~ 30 µm) originate from the Bragg resonance. The distinction between the two effects results primarily from different distributions of orientations of the bricks and the relative contribution of these orientations to scattering. In order for a particular photonic brick to display strong iridescence, it must be oriented such that the angle between its surface and the observer ($\theta_{obs}$) matches the angle between its surface and the light source ($\theta_{illum}$) in a specular arrangement. The peak wavelength of this iridescence blueshifts as the angle between the source and detector ($\theta_{obs} + \theta_{illum}$) increases. When the photonic bricks are dispersed over a wide range of orientations, all possible pairings of $\theta_{obs}$ and a $\theta_{illum}$ will have roughly equal small fractions of the photonic bricks that display the bright iridescence. This leads to a sparkly appearance, such as in Fig. 7-iv, where at any angle a small number of photonic bricks are brightly iridescent, while the majority display the red hue produced by the AuNS in combination with any residual non-resonant scattering (e.g. scattering from surfaces and defects, multiple scattering, etc.). When the photonic bricks display a preferential orientation, as they do with the substrate in Fig. 7A-iii, their iridescence resembles more closely that of a flat film, appearing as a more of a homogeneous color travel when the light source and observer locations allow for specular reflections from the plane of preferred orientation.

Fig. 8A shows variable-angle spectra, taken under specular illumination (left column) and with fixed illumination at normal incidence with respect to the substrate (right column), for photonic bricks with different overall dimensions, but having the same porosity ($a$ = 350 nm, monodispersed) and AuNS concentration (0.7% by matrix volume, as in Fig. 7A iii-vi). As do the images in Fig. 7A, these spectra illustrate the trend going from sparkle to color travel with increasing width. The specular reflection spectra for the freeform and 25 µm x 100 µm photonic bricks are dominated by the Bragg resonance (from bricks oriented parallel to the substrate), which blueshifts with angle. However, this resonance is much less prominent for the 25 µm x 50 µm photonic bricks and is only faintly visible over top of considerable background scattering whose only wavelength selectivity is an angle-independent suppression of scattering in the AuNS absorption band. The spectra at normal-incidence illumination of the 25 µm x 50 µm photonic bricks show this prominent background as well, however sparkle is also evident from several bright and narrow scattering peaks that can be seen at a few angles (e.g., 25°, –45°). These bright spots were not observed at the same angles in different samples of the same dimensions. They occurred sporadically wherever the detector happened to catch one or more of the photonic bricks at the correct orientation to display a strong Bragg reflection. In contrast, scattering from normal-incidence illumination is highly concentrated at small angles and over a narrow wavelength range for the 25 µm x 100 µm and freeform photonic bricks, indicating that many of them are preferentially orienting with the substrate. Fig. 8B quantifies the angular distribution of scattering from normal-incidence illumination, summed over three samples for each set of dimensions.

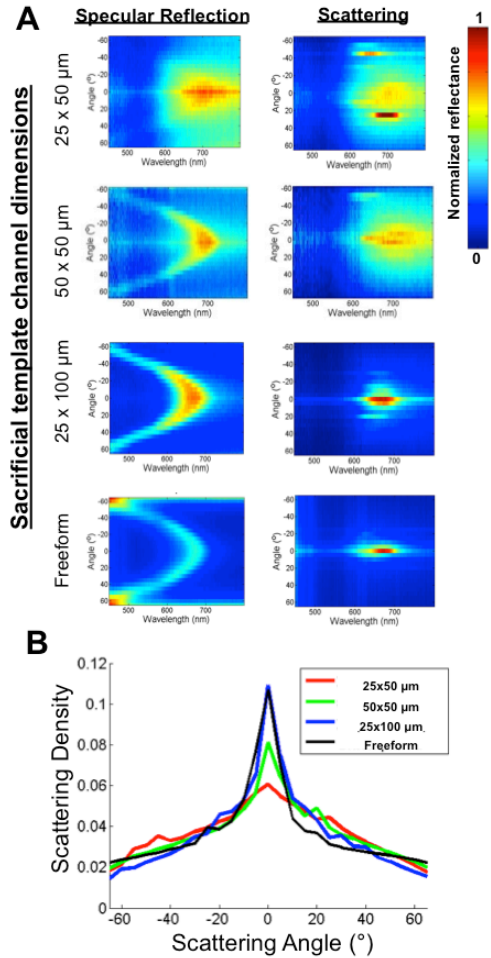

Fig. 8. (A) Representative variable-angle spectra of photonic bricks with different dimensions (height x width). All have 0.7% AuNS by matrix volume and ordered porosity ($a = 350$ nm). Photonic bricks were mounted on a flat substrate and spectra were taken with specular illumination (left column) and normal-incidence illumination (right column) with respect to the orientation of the substrate. (B) Angular distribution of scattering at normal incidence illumination (averaged over all wavelengths and 3 samples of each type). The full widths at half-maximum of these angular scattering distributions are: 95° (25 μm x 50 μm), 40° (50 μm x 50 μm), 15° (25 μm x 100 μm), and 14° (freeform).

In order to be able to impose a preferred orientation onto the photonic bricks, they must have sufficiently asymmetric overall dimensions. When such an asymmetry exists, there are many ways it can be exploited to orient the bricks to produce color travel when the pigment is deposited onto a surface, either dry or in a paint formulation. In our case, the light pressure applied to adhere the bricks to the tape on the substrate is what creates the preferred orientation parallel to the substrate. Although all of our photonic bricks have a length that far exceeds their other dimensions (see Fig. 2), preferred orientation will not be induced unless their width is also significantly greater than their height. Therefore, only the 25 μm x 100 μm and freeform (<w> ~ 110 μm and <h> ~ 30 μm) photonic bricks, both with $w/h$ aspect ratios greater than 3, show easier preferential alignment than the 25 μm x 50 μm and 50 μm x 50 μm bricks that have smaller aspect ratios. The ability to engineer asymmetric shapes is an important degree of freedom that allows control of macroscopic visual effects, in this case letting us choose between sparkle and color travel. Controlling the overall size of the photonic bricks will also affect the texture of the pigment's appearance and whether or not individual particles can be resolved by eye (see for example Fig. 7A-iii-vi, above).

## 5. Discussion

The hierarchical structure of the photonic bricks produced by this templated co-assembly procedure provides many easily accessible tuning degrees of freedom in the morphology across several disparate length-scales. These tuning parameters enable complex visual properties of the pigment to be engineered by allowing the absorption profile and different aspects of the scattering profile to be independently designed. Fig. 9 shows an image of Olympic rings formed by several different types of freeform photonic bricks under diffuse illumination and at different angles under specular illumination. This picture illustrates the large variety of appearances that can be produced through manipulation of the plasmonic absorbers and the porosity within this co-assembly method. The three rings in the top row (i, iii, v) contain plasmonic absorbers and exhibit these hues. The different hues are produced by choosing different plasmonic absorbers (i – blue from Ag nanoplates; iii,v – red from AuNS). The intensity of the hue is tunable via the concentration of the plasmonic absorber. This effect can be seen by comparing the dark red ring (iii, 2.4% AuNS by matrix volume) and the bright red ring (v, 0.7% AuNS by matrix volume).

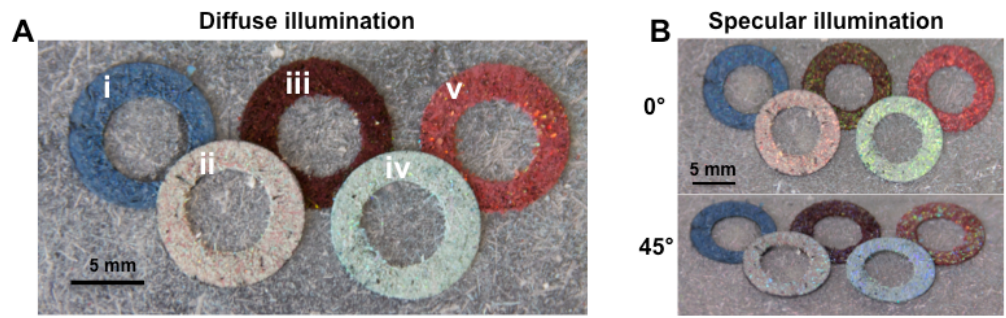

Fig. 9. Images of the Olympic rings formed from different types of photonic bricks, illustrating different combinations of hues and effects, imaged from normal incidence under diffuse illumination (A) and at 0° and 45° under specular illumination (B): Background is provided by the freeform photonic bricks with disordered bidispersed porosity ($a_1$ = 240 nm, $a_2$ = 350 nm) containing no plasmonic absorber and appearing as matte white with no significant sparkle or color travel in all images; (i) – freeform photonic bricks with ordered porosity ($a$ = 240 nm) and containing 0.35% (by matrix volume) Ag nanoplates ($\lambda_{abs}$ ~ 650 nm), appearing blue from all angles with a blue sparkle also observable in specular reflection; (ii) – a mixture of freeform photonic bricks, all having ordered porosity ($a$ = 350 nm), half of which contain 0.15% AuNS (by matrix volume) and the other half - no plasmonic absorber, appearing pale yellow under diffuse illumination with more of a red shine under specular illumination at 0°, which blueshifts to a blue-green at 45°; (iii) – freeform photonic bricks with ordered porosity ($a$ = 300 nm) containing 2.4% AuNS (by matrix volume), displaying a deep red hue along with a green iridescence at 0° which becomes blue at 45°; (iv) – freeform photonic bricks with ordered porosity ($a$ = 300 nm) containing no plasmonic absorber, appearing pale green under diffuse illumination and having a strong green iridescence at 0°, which becomes blue at 45°; (v) – freeform photonic bricks with ordered porosity ($a$ = 350 nm) containing 0.7% AuNS (by matrix volume), displaying a bright red hue along with a red iridescence at 0°, which becomes green at 45°.

The porosity of the photonic bricks – tuned via the size and monodispersity of the polymer colloids – provides scattering that can enhance the brightness of the hue and/or exhibit iridescence. Disordered porosity (from polydispersed polymer colloids) offers an angle-independent appearance and is used to produce the matte white of the background in Fig. 9. Ordered porosity leads to strong Bragg resonances. The resonance wavelength is tunable by the size of the polymer colloids, and can be chosen to enhance the hue provided by the plasmonic absorbers, as is shown in rings i ($a$ = 240 nm) and v ($a$ = 350 nm), or contrast with this hue, resulting in a composite appearance. This is shown in ring iii, where a deep red hue from AuNS doping contrasts with a green iridescence at normal incidence provided by the porosity ($a$ = 300 nm). These Bragg resonances are also angle-dependent, blueshifting with increasing viewing angle under specular illumination, and providing a prominent color travel effect in freeform photonic bricks (see Fig. 7A). This produces a composite appearance when plasmonic absorbers are present (e.g. Fig. 9, i,ii,iii,v), combining an angle-independent hue, with an angle-dependent iridescence displaying color travel. As a result, the hue and iridescence can produce the same color at some angles (e.g. Fig. 9, i,v at normal incidence) or contrasting colors (e.g. Fig. 9, iii at normal incidence; iii,v at 45°). This color travel, providing a prominent angle-dependent color that contrasts with the underlying angle-independent hue from the plasmonic absorber, could be transformed into more of a diffuse sparkle by fixing the height and width of the photonic bricks into more symmetric shapes (e.g. as shown in Fig. 7A, iii, iv).

In the presence of little (Fig. 9, ii) or no (Fig. 9, iv) plasmonic absorber, the Bragg resonance from the ordered porosity provides the dominant source of color, and the appearance displays more of a uniform color travel with no contrasting, angle-independent hue. For example, ring iv (Fig. 9) under specular illumination appears uniformly green at normal incidence and uniformly blue at 45°. However, the absence of a strong absorption from little (Fig. 9, ii) or no (Fig. 9, iv) plasmonic absorber leads to very pale colors under diffuse illumination. If one wanted to enhance the saturation of this angle-dependent, iridescent color without introducing a second distinct, angle-independent hue, a broadband absorption profile would be desired. This could be accomplished within the framework of our synthetic protocol by doping the photonic bricks with either a single broadband absorber such as carbon black or with a mixture of plasmonic absorbers with different sizes and shapes.

The simple one-pot synthesis method described here enables production of pigment particles that display a large variety of colors and effects all based on a small set of chemically stable, robust, and non-toxic inorganic materials. This approach has potential to expand and simplify the development of new colorants, particularly in those markets where long-term stability and low toxicity have high priority. This method also introduces some trade-offs that merit future exploration. These include the trade-off between the enhanced color with added design freedom provided by plasmonic absorbers and the higher cost of good plasmonic materials (e.g. gold) compared to a broadband absorber such as carbon black.

A second potential drawback of this method is the use of silica as the matrix material. While silica is a highly abundant and inert dielectric material, it has a fairly low refractive index ($n \sim 1.46$) compared to other dielectrics such as titania that are commonly used in colorants. With a decreased refractive index contrast comes weaker scattering at each interface, which requires our photonic bricks to be larger in size (e.g. contain more periods of the structure) than those of higher index in order to produce the same effects. On the other hand, lower refractive-index contrast also leads to narrower Bragg resonances and therefore purer colors.

The low refractive index also requires us to have stable air-filled porosity to obtain strong scattering, potentially posing difficulties when incorporating the photonic bricks into liquid paint formulations. However, with the appropriate surface functionalization, the inverse-opal geometry of the photonic bricks can resist infiltration of liquids when submerged, maintaining their color when dispersed in liquid paint formulations. Due to the highly re-entrant geometry of the inverse-opal pore structure,[43-44] infiltration of the pores by a liquid only occurs for very small intrinsic contact angles ($\theta_c < \sim 20°$). As a result, when functionalized with strongly hydrophobic perfluorinated surface groups, pore filling can be prevented for nearly any liquid, including low-surface-tension oils such as those commonly found in paints.[44-45] Photonic bricks were rendered impermeable to most liquids by functionalizing their surfaces with perfluorinated surface groups (shown in Fig. 10A,B), through exposure to vapors of 1H,1H,2H,2H-tridecafluorooctyl trichlorosilane (13FS) after synthesis. In contrast to non-functionalized photonic bricks, 13FS-functionalized bricks maintain their appearance after immersion in mineral oil (Fig. 10A) and epoxy resin (Fig. 10B, UVO-114, imaged after UV curing).

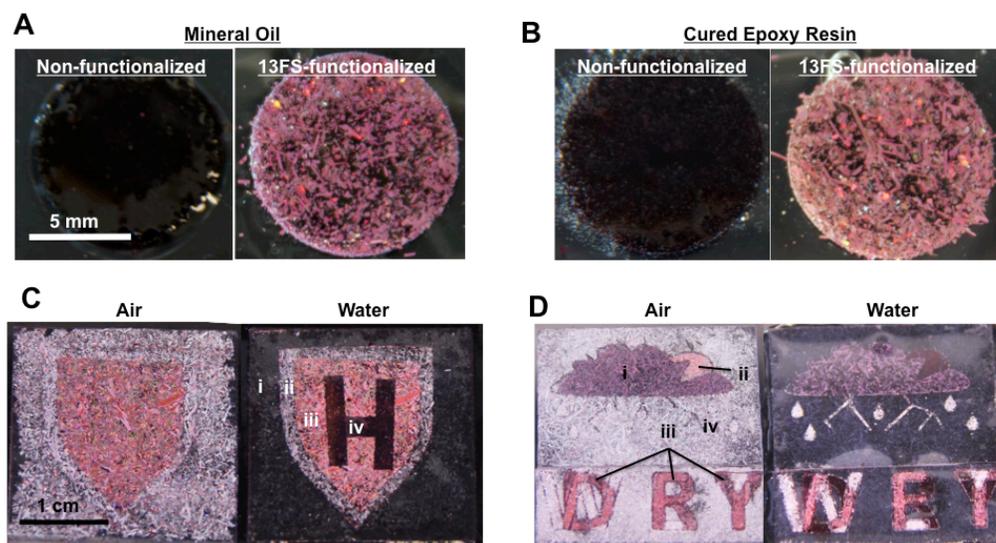

Fig. 10. (A,B) Maintaining color in formulation. Two sets of photographs (diffuse illumination) comparing non-functionalized photonic bricks (left in each set) with those functionalized with 1H,1H,2H,2H-tridecafluorooctyl trichlorosilane (13FS) (right in each set) when immersed in mineral oil (A) and UV-curable epoxy resin (B, UVO-114, imaged following UV-curing). All four samples are freeform photonic bricks with $a$ = 350 nm and 0.7% AuNS content (by matrix volume). This functionalization allows the air porosity to be maintained in liquid formulations, critical for maintaining their optical appearance. (C,D) Adaptive paints. Photographs (diffuse illumination) of freeform photonic bricks (all with $a$ = 350 nm) painted onto a surface with patterns of color and patterned surface chemistry, revealing different images when wet and dry. (C) Pattern of color: i,ii – Freeform photonic bricks with no plasmonic absorber, iii, iv – freeform with 0.7% AuNS content (by matrix volume); Pattern of surface functionalization: i,iv - not functionalized, ii,iii – 13FS-functionalized. When immersed in water, brightness of color dramatically diminishes in only the non-functionalized regions where the air pores have become filled. (D) A more complex color pattern produced by varying AuNS content (i – 1.3%, ii – 0.4%, iii – 0.7%, iv – 0%) showing the same region-selective color change in water. Color is retained in regions that were selectively functionalized with 13FS (all of region i, and the portions of regions iii and iv that remain bright in water).

The surface modification changes the properties of the photonic bricks on a length-scale that is smaller than the plasmonic particles and visible wavelengths, and therefore do not affect the appearance in the absence of a liquid. Thus patterns of surface functionality can be used to create dynamic images that contain encrypted patterns of surface chemistry. These encrypted patterns become apparent only when wet.[44-46] Subsequent liquid exposure reveals pre-determined patterns based on where the liquid is able to penetrate the pores, due to the reduced refractive index contrast between the silica matrix and the infiltrating liquid. This "dynamic paint" effect is illustrated in Fig. 10C,D. In both examples, images have been created using freeform photonic bricks with differing AuNS content. Although not visible when dry, the surface functionalizations form distinct spatial patterns that are revealed when wet. This feature, resulting from the open porosity of the pigment particles, allows images formed from differently functionalized pigments to adapt to their environment, enabling applications such as weather-adaptive road signs.

## 6. Conclusions

In summary, we have presented a simple one-pot co-assembly method for synthesis of a structural color pigment consisting of silica inverse-opal photonic bricks doped with nanoscale plasmonic absorbers. The structure and composition of the photonic bricks can be independently tuned across several different length-scales, from the submillimeter scale shape of the bricks, to their wavelength-scale porosity, to the nanometer scale plasmonic absorbers, to molecular-scale surface functionalization. Spectrally selective absorption from the nanoscale plasmonic absorbers provides an angle-independent hue that is tunable by adjusting their size, shape and concentration. The wavelength-scale pore geometry tunes the

wavelength- and angle-dependence of the scattering profile, and can be engineered to produce angle-dependent Bragg resonances that can either enhance or contrast with the plasmonic color. Manipulating the symmetry in the overall dimensions of the photonic bricks allows their preferential alignment in an environment with broken symmetry to be either enabled or suppressed, leading to color travel or sparkle respectively. By modifying their surface chemistry, we can further engineer interactions between liquids and the pores. This allows the photonic bricks to maintain their colorful dry appearance when mixed into oil-based liquid formulations, by exploiting the highly re-entrant geometry of the inverse-opal pore structure. Spatial patterning of surface chemistry also enables the creation of dynamic images using the photonic bricks, which display a selective, liquid-responsive appearance. The freedom to engineer a multitude of complex and dynamic optical appearances within a single inorganic material framework is enabled by the ability to independently adjust the morphology of the photonic bricks across several disparate length-scales.

## Appendix A. Methods

### A1. Fabrication of photonic bricks

The methods required for inverse opal fabrication and functionalization have been previously reported[33,37,45-46] and were adapted to produce inverse opal photonic bricks. These bricks are categorized as "freeform" (denoting the absence of a lateral template during co-assembly) and "templated" (denoting the presence of a photoresist template with channels during co-assembly).

### A1.1 Preparation of plasmonic absorbers

An aqueous solution of silica-coated silver nanoplates (1 mg/mL Ag content, 60-80 nm diameter and a 10 nm thickness, $\lambda_{max} \sim 650$ nm) was purchased from Nanocomposix, Inc. and used in the co-assembly precursor solution without further purification.

A solution of citrate-stabilized gold nanospheres (AuNS) was prepared according to a modified literature procedure.[40] Sodium citrate (50 mg, 0.17 mmol) was added to a refluxing, vigorously stirred solution of sodium tetrachloroaurate dihydrate (25 mg, 0.065 mmol) in triple DI water (123 mL). The mixture was stirred under reflux for 15 min before it was allowed to cool to room temperature. The initially yellow solution changed to deep red and was stored in a glass bottle protected from light. The AuNS formation was confirmed by UV-Vis spectroscopy ($\lambda_{max} \sim 520$ nm in water), and by TEM ($d \sim 12$ nm). The molar concentration of the AuNS solution was calculated from the UV-Vis spectrum, using a method previously described in the literature.[47]

Subsequently, methoxy-poly(ethylene glycol)-thiol (mPEG-SH, Mw = 5000) was covalently grafted to the surface of the AuNS by dropwise addition of 2 ml of mPEG-SH (25 mg/mL) solution to the vigorously stirred as-prepared AuNS solution.[48-49] The solution was stirred for ~2 h allowing citrate ligands to exchange with mPEG-SH. PEG-modified particles were then centrifuged (14000 rpm, 1h) twice to remove excess of mPEG-SH and concentrated to 12 ml to give ~160 nM Au-PEG solution.

The approximate amount of AuNS in the as prepared solution was ~$1.4 \times 10^{15}$ NS and the concentration was ~16 nM. The amount of AuNS incorporated into the photonic bricks during the co-assembly process was determined by digesting 30 mg aliquots of photonic bricks (combined from several repeats of the fabrication protocol with the same amount of AuNS solution added) in an acid solution and analyzing by ICP (inductively coupled plasma – atomic emission spectroscopy, conducted by Massachusetts Materials Research Inc.). The acid digestion procedure was conducted as follows: Each 30 mg aliquot was first immersed in 1 mL of 50% hydrofluoric acid at 60°C to digest the silica. Once the silica was completely digested, 1 mL of nitric acid and 3 mL of hydrochloric acid were added to digest the AuNS. This mixture was heated at 120°C for ~3 h until the liquid had reduced to a total volume of approximately 1 mL. The sample was then reconstituted in DI water to a total mass of 50 g before being sent out for ICP analysis.

### A1.2. Preparation of Photoresist Template

Glass microscope slides were cleaned in acid piranha (1:3 $H_2SO_4$:30% $H_2O_2$) for at least 20 min, followed by exposure to plasma oxygen for 5 min. Slides were kept at 180°C on a hot-plate and naturally cooled to room temperature immediately prior to mask photolithography. A base sacrificial layer of SU-8 2010 (Microchem) was spun onto the slide before flood exposure of UV light (365 nm), for a final thickness of 10 µm. After a post-exposure hard-bake (95°C), a secondary layer of SU8 was deposited. SU-8 2025 and SU-8 2050 were used to fabricate channels with a height of 25 µm and 50 µm respectively. After a soft (65°C) and hard (95°C) bake, slides were masked with Mylar masks (FineLine Imaging) during flood exposure of UV light (365 nm). After post-exposure soft and hard bake, slides were submerged in SU-8 developer (Microchem) until sufficiently developed.

### A1.3. Co-assembly

The co-assembly precursor solution typically comprised of three elements suspended in water: 1:1.5:1 ratio of 0.01M HCl:ethanol:TEOS (tetraethylorthosilicate) solution (aged for 1 hour), an aqueous polymer colloids, made either of polymethylmethacrylate (PMMA, for $a$ = 300 nm, 350 nm) or polystyrene (PS, for $a$ = 240 nm) (~6.75 wt%) and an aqueous solution of plasmonic absorbers (either AuNS or Ag nanoplates). Water was added as needed to achieve the desired relative and overall concentrations of the precursors.

Prepared glass slides with SU-8 channels were cleaned via oxygen plasma for 5 min to induce hydrophilicity before vertical suspension in 20 mL scintillation vials containing the co-assembly solution in an oven (Memmert) at 65°C. Typical time for complete evaporation was 48 h. Slides were placed in a furnace that was ramped from room temperature to 500°C over 5 h and held at that temperature for 2 h. This step served to sinter the matrix, remove the polymer colloids, and release the photonic bricks from the photoresist template. Typical yields of templated photonic bricks were 3-10 mg per slide.

Freeform were derived from identical solutions placed in 20 mL scintillation vials at 65°C for 72 h. Formation occurred spontaneously due to natural cracking of the film deposited on the wall, facilitating particulate release for collection during sintering. Since the Ag nanoplates were found to degrade at high temperatures, solvent extraction (using toluene) was used to remove the polymer colloids instead of high-temperature calcination for the freeform photonic bricks doped with Ag platelets (Fig. 3B, Fig. 7-i).

### A1.4. Surface functionalization

Photonic bricks stored in a 20 ml scintillation vial were plasma-cleaned for 10 min and then placed in a low-volume, low-pressure vacuum desiccator for 24 h with a small vial of 1H,1H,2H,2H- tridecafluorooctyltrichlorosilane. Upon completion, the vial was placed on a hot plate at 150°C for 15 min before handling. To create patterns of photonic bricks with different surface chemistry, laser cut paper masks were used to cover selected areas of a large (1x1 inch) square of double-stick adhesive (3M, Inc.). Photonic bricks were carefully placed and pressed against adhesive areas before the paper masks are removed. Subsequent iterations are done until the entire surface area was covered.

### A2. Characterization

Photonic bricks were deposited onto glass substrates coated with double-sided tape for imaging and spectral analysis. The tape-coated substrates were gently pressed against the inside of glass vials containing the photonic bricks. Once the entire substrate was covered with photonic bricks, those that were not adhered to the tape were removed by gently pressing the substrate with a lint-free wipe. Optical images of samples were acquired using a Canon EOS 60D digital camera equipped with a macro lens and mounted on a tripod. A small lamp was fixed at an approximate distance of 1', illuminating the sample at the same angle as collection for images of specular scattering (Fig. 7 - middle and bottom rows, Fig. 9B). For diffuse illumination (Fig. 5C, Fig. 7 – top row, Fig. 9A, Fig. 10), the camera was mounted at normal incidence to the sample. The sample was illuminated from one side by a wide desk lamp ($d \sim 8$"), placed roughly 1' away from the sample with a sheet of paper placed in between the lamp and the sample, which served as a diffuser for the incident light. All photographs (both specular and diffuse illumination) were taken in the same spot on the lab bench. Physical characterization of photonic bricks dimensions and porosity and confirmation of metal nanoparticle incorporation were provided by: i) SEM images of photonic bricks and SU-8 channels (Zeiss Ultra Plus and Supra55VP); and ii) High-magnification TEM images of the inverse opal matrix doped with plasmonic nanoparticles (JEOL 2100). Photonic bricks dimensions were measured from SEM images using ImageJ.

Reflectance spectra for IOF areas that contained a fixed and well-defined number of close-packed layers (Fig. 6) were collected using a custom-built optical microscope (Leica). Variable angle spectra (specular reflectance and diffuse scatter) of photonic bricks were obtained using a custom-built variable angle spectrometer.[35] Specular reflection was measured by illuminating the sample at angles ranging from $-65°$ to $65°$ (angles listed with respect to the substrate) in steps of $5°$ and collecting spectra at corresponding angles across the normal. Diffuse scattering was measured by illuminating the sample at 0°C and setting the detector to rotate around the sample in an angle range of $-65°$ to $65°$ in steps of $5°$ and at a higher resolution from an angle range of $-30°$ to $30°$ in steps of $1°$. Each sample was mounted onto black double stick adhesive and spot sizes were held consistent at ~3 mm in diameter. Each measurement was repeated three times per sample. Coverage of photonic bricks on the substrate was ensured to be of sufficient density that the underlying substrate was not visible to the eye, and the measured spectra were reproducible and did not contain artefacts of the substrate (which appeared as large, broadband signals in specular reflection).

**A3. Modeling Absorber Effects in Inverse Opal Films**

A simplified optical model was used to compare the effects of doping inverse-opal films (IOFs) with AuNS (spectrally selective absorber) against the incorporation of an ideal broadband absorber that has a wavelength-independent absorption profile. The results are shown in Fig. 6B and are compared with experimental results shown for IOFs with AuNS and with no absorber. To estimate the reflectance, the inverse-opal structure was approximated as a 1D multilayer by taking the area-averaged dielectric constant in the lateral dimensions at each value of the *z*-coordinate. The reflectance spectrum of this structure was then calculated using the transfer matrix method. The wavelength-dependent optical constants ($n$ and $k$) of the AuNS-doped silica composite (1% AuNS by volume vs. the matrix) were calculated using the effective medium model described in Ref. 42. For an ideal broadband absorber, a wavelength-independent $k$ was set at 0.008, which gave similar absorption to the AuNS in the blue and green wavelengths.

**Acknowledgements**

The authors thank Dr. Caitlin Howell, Derek Cranshaw and Charlie Payne for helpful discussions. The work was supported by the US AFOSR under award number FA9550-09-1-0669-DOD35CAP and in part by the BASF SE. Template microfabrication and electron microscopy of the photonic bricks were performed at the Center for Nanoscale Systems (CNS) at Harvard University, a member of the National Nanotechnology Infrastructure Network (NNIN), which is supported by the NSF under award number ECS-0335765. M.K. acknowledges the financial support from the Alexander von Humboldt Foundation in form of a Feodor Lynen postdoctoral research fellowship. K.R.P. acknowledges support from a DoD National Defense Science and Engineering Graduate Fellowship. N.V. acknowledges support from the Leopoldina fellowship.